\documentclass[aps,prd,nofootinbib,twocolumn,superscriptaddress,preprintnumbers]{revtex4-1}

\usepackage{amssymb}
\usepackage{amsmath}
\usepackage{epsfig}
\usepackage{breakurl}
\usepackage{bm}
\usepackage{xcolor}
\usepackage{tikz}
\usetikzlibrary{decorations.markings}

\usepackage{subfigure}

\makeatletter
\def\simgt{\mathrel{\lower2.5pt\vbox{\lineskip=0pt\baselineskip=0pt
           \hbox{$>$}\hbox{$\sim$}}}}
\def\simlt{\mathrel{\lower2.5pt\vbox{\lineskip=0pt\baselineskip=0pt
           \hbox{$<$}\hbox{$\sim$}}}}
\makeatother

\newcommand{\bel}{\begin{align}}
\newcommand{\eel}{\end{align}}

\newcommand{\Eq}[1]{Eq.~\ref{#1}}
\newcommand{\Eqs}[2]{Eqs.~\ref{#1} and \ref{#2}}

\newcommand{\vev}[1]{\langle #1 \rangle}
\newcommand{\bra}[1]{\langle #1 |}
\newcommand{\ket}[1]{| #1 \rangle}

\newcommand{\nn}{\nonumber}

\newcommand{\beq}{\begin{equation}}
\newcommand{\eeq}{\end{equation}}
\newcommand{\bea}{\begin{eqnarray}}
\newcommand{\eea}{\end{eqnarray}}
\newcommand{\mo}{\mathcal{O}}

\newcommand{\eps}{\epsilon}

\def\topbotatom#1{\hbox{\hbox to 0pt{$#1\bot$\hss}$#1\top$}}

\begin{document}

\title{Manifestly Causal In-In Perturbation Theory about the Interacting Vacuum}

\author{Matthew Baumgart}
\affiliation{Department of Physics, Arizona State University, Tempe, AZ 85287, USA}
\author{Raman Sundrum}
\affiliation{Maryland Center for Fundamental Physics, University of Maryland, College Park, MD 20742}

\begin{abstract}

In-In perturbation theory is a vital tool for cosmology and nonequilibrium physics.  Here, we reconcile  an apparent conflict between two of its important aspects with particular relevance to  De Sitter/inflationary contexts: (i) the need to slightly deform unitary time evolution with an $i\eps$ prescription that projects the free (``Bunch-Davies") vacuum onto the interacting vacuum and renders vertex integrals well-defined, and (ii) Weinberg's ``nested commutator'' reformulation of in-in perturbation theory which makes manifest the constraints of causality within expectation values of local operators, assuming exact unitarity. We show that a modified $i\eps$ prescription maintains the exact unitarity on which the derivation of (ii)  rests, while nontrivially agreeing  with (i) to all orders of perturbation theory.

\end{abstract}

\maketitle

\section{Introduction}

The commutator plays a crucial role in relativistic quantum field theories.  It precisely captures the causal propagation of physical effects.  Cosmic Microwave Background (CMB) observation is an archetypical experiment sensitive to these causality constraints.  Any correlation recorded on it must have arisen in the past lightcone of the correlation points.  Predictions for the primordial correlations seen in CMB probes are typically computed with in-in perturbation theory, as we are measuring expectation values within a state at a particular time.\footnote{This technique is also referred to as Closed Time Path (CTP) or Schwinger-Keldysh.}  
Often, for practical purposes, it is sufficient that deep principles are ``buried somewhere in the formalism" as long as the method of calculation is efficient.
  Nonetheless, formulations that make fundamental principles and symmetries manifest often bring a sharpening clarity to conceptual questions and can motivate new approximations.
  In a recent work by the present authors making heavy use of explicit causality \cite{Baumgart:2019clc}, we showed how the nonperturbative reorganization of infrared divergences in light scalar De Sitter field theory, posited originally by Starobinsky as ``stochastic inflation'' \cite{Starobinsky:1986fx}, arises efficiently from considerations of power counting and the all-orders structure of in-in perturbative diagrams. Weinberg's ``nested commutator" reformulation of  in-in perturbation theory, with its more explicit rendering of causality constraints, played a central role in our analysis. 

The standard approach to in-in perturbation theory computes expectation values in the interaction picture, with ket and bra  versions of the same state multiplying the time evolution operator, $U$, given by the interacting hamiltonian $H_I$,  and its hermitian conjugate, $U^\dag$:
\begin{align}
\vev{\mo}(t) &= \bra{\Omega} \left( \overline T e^{+i \int_{-\infty}^t H_I dt^\prime} \right) \mo \left(T e^{-i \int_{-\infty}^t H_I dt^\prime} \right) \ket{\Omega}.
\label{eq:tradinin}
\end{align}
In \cite{Weinberg:2005vy} though, Weinberg sketched an argument for an equivalent, alternative determination of the time-dependent expectation value:
\begin{align}
\vev{\mo}(t) &= \sum_{V=0}^\infty (-i)^V \int_{- \infty}^t dt_V \ldots \int_{-\infty}^{t_3} dt_2 \int_{-\infty}^{t_2} dt_1 \nn \\
&\times  \Big \langle \Omega \Big | \Big [ \Big [ \ldots \Big [ \mo,  H_I(t_V) \Big ] \ldots , H_I(t_2) \Big ], H_I(t_1) \Big ] \Big | \Omega \Big \rangle.
\label{eq:pertex}
\end{align}
We fleshed the argument out into a complete formal derivation of Eq.~\ref{eq:pertex} in the Appendix of \cite{Baumgart:2019clc}.  In particular, the presence of the nested commutators makes causality manifest.  This, combined with the development of a diagrammatic formalism by Ref.~\cite{Musso:2006pt} to make operational use of Eq.~\ref{eq:pertex} allowed us to give a clear, rigorous graphical proof of stochastic inflation, heretofore elusive for 30+ years.  See Refs. \cite{Gorbenko:2019rza,Mirbabayi:2019qtx,Cohen:2020php} for alternative recent discussions.

We wish to address here though, a technical detail missing from the demonstration of the equivalence of \Eqs{eq:tradinin}{eq:pertex}.  The proof we gave in \cite{Baumgart:2019clc} is straightforward starting from Eq.~\ref{eq:tradinin} as given, and even holds at the operator level, dropping the states from the RHS of it and Eq.~\ref{eq:pertex}.  At the mathematical level, one is just reorganizing the operator algebra.  For any {\it given} $\ket{\Omega}$, one can thus compute the expectation value either way.  In practice though, doing time-dependent perturbation theory typically means we do {\it not} have an explicit external state to compute with.  Typically, we seek to calculate expectation values of the vacuum state of the {\it interacting} theory (which we continue to call $\ket{\Omega}$) in terms of the free-theory ground state, $\ket{0}$.  Thus, in order to account for the nontrivial projection of the latter  onto the former, similarly to  in-out perturbation theory, the trick is to  deform Eq.~\ref{eq:tradinin} to 
\begin{align}
\vev{\mo}(t) &= \frac{\bra{0} \left( \overline T e^{+i \int_{-\infty(1+i\eps)}^t H_I dt^\prime} \right) \mo(t) \left(T e^{-i \int_{-\infty(1-i\eps)}^t H_I dt^\prime} \right) \ket{0}}
{\bra{0} \left( \overline T e^{+i \int_{-\infty(1+i\eps)}^t H_I dt^\prime} \right) \left(T e^{-i \int_{-\infty(1-i\eps)}^t H_I dt^\prime} \right) \ket{0}}.
\label{eq:inindef}
\end{align}
As pointed out in Refs.~\cite{Adshead:2008gk,Senatore:2009cf}, this would seem to spoil Weinberg's Eq.~\ref{eq:pertex} as an equivalent computation.  In particular, the $\eps$ shifts into the complex plane are different for bra and ket evolution, and we cannot blithely interchange terms coming from $U_{\eps}^\dag$ and $U_{\eps}$, as the commutators would have us do.  Furthermore, $U_{\eps}^\dag$ and $U_{\eps}$ are no longer unitary, and thus not inverses of each other, leading to the nontrivial denominator, which gives the familiar (from in-out perturbation theory) division by ``vacuum bubbles" in order to account for the projection factor from $|0\rangle$  onto $| \Omega \rangle$.

While the roles played by $\eps$ are crucial in enabling projection onto the interacting vacuum and rendering perturbative vertex integrals to $t \rightarrow - \infty$ well-defined, 
the detailed form of the $\eps$-prescription  is quite flexible at the perturbative level.
Ref.~\cite{Kaya:2018jdo} recognized this flexibility and proposed an alternative $\eps$ deformation that would maintain exact unitarity, and hence the derivation of a nested commutator formulation of in-in perturbation theory.  It claimed perturbative equivalence of its new $\eps$ prescription to Eq.~\ref{eq:inindef}. 
This is  a valuable insight, but the argument given in  \cite{Kaya:2018jdo} for this perturbative equivalence  has an important loophole. 

Indeed, such a loophole had to exist because otherwise the argument of  \cite{Kaya:2018jdo} would have shown the equivalence of just the numerator of Eq.~\ref{eq:inindef} to a nested commutator form, missing the subtlety of vacuum bubbles.  Here, we present a complete treatment, showing that in-in perturbation theory can be made manifestly unitary at all intermediate steps, as well as manifestly causal.  In particular, we show the perturbative equivalence of Eq.~\ref{eq:inindef} to the nested commutator form
\begin{align}
&~ \vev{\mo}(t) = \sum_{V=0}^\infty (-i)^V \int_{- \infty}^t dt_V \ldots \int_{-\infty}^{t_3} dt_2 \int_{-\infty}^{t_2} dt_1 \nn \\
&\times  \Big \langle 0 \Big | \Big [ \Big [ \ldots \Big [ \mo,  H^{\eps}_I(t_V) \Big ] \ldots , H^{\eps}_I(t_2)  \Big ], H^{\eps}_I(t_1) \Big ] \Big |  0 \Big \rangle,
\label{eq:pertexprime}
\end{align}
where 
\begin{equation}
H^{\eps}_I(t) \equiv H_I e^{\eps t}
\end{equation}
is clearly hermitian.
Note that in  Eq.~\ref{eq:pertexprime} the state is just the free vacuum $|  0 \rangle$, while the interaction Hamiltonian  is multiplied by an $\eps$ damping factor, giving rise to a well-defined perturbative expansion that can be compared with that of Eq.~\ref{eq:inindef}.

In Section \ref{sec:mink}, we show the equivalence of Eq.~\ref{eq:inindef} and Eq.\ref{eq:pertexprime} for the easier case of Minkowski space, before extending the argument to De Sitter in Section \ref{sec:ds}.  It would be interesting to further investigate if the same nested commutator reformulation can be derived more generally for in-in perturbation theory in any past-eternal background spacetime, but this lies beyond the scope of this paper.

\section{Minkowski Space}
\label{sec:mink}

We begin with the technically simpler argument for flat-space in-in perturbation theory.  The core of the argument will be the same for our cosmological case of interest, but this warmup has the advantage of very simple calculus.  Ultimately, we will demonstrate that the reorganization of in-in perturbation theory due to Ref.~\cite{Weinberg:2005vy} is legitimate, even in the presence of $\epsilon$-deformed contours that extend to $t \rightarrow -\infty$ to project onto the full interacting vacuum from the free theory, or Fock, vacuum.  To do so though, we begin with the conventional in-in formulation and its standard $i\eps$ deformation, \Eqs{eq:tradinin}{eq:inindef}, respectively.

The reason why in-in perturbation theory is ``richer'' than familiar in-out is the presence of the two different quantum mechanical evolution operators, $U(\bar{t}) \equiv T e^{-i \int_{-\infty}^{\bar{t}} H_I dt}$ and $U^\dag(\bar{t})$. We reserve $\bar{t}$ to denote the correlation time from here on for clarity.  Upon deformation, these are no longer inverses of each other, which is why the denominator in Eq.~\ref{eq:inindef} is nontrivial, and this violation of unitarity precludes the naive use of Weinberg's reformulation.  In in-in perturbation theory, we typically work with four different propagators, depending on whether the fields arise from $U,\, U^\dag$ or a mix.  We explicitly treat only scalars, writing $\phi^+$ for those from $U$ and $\phi^-$ for those from $U^\dag$, and thus have
\begin{align}
G^{++}(t,t^\prime) &= \vev{T\, \phi^+(t) \phi^+(t^\prime)} \nn \\ 
G^{--}(t,t^\prime) &= \vev{\overline T\, \phi^-(t) \phi^-(t^\prime)} \nn \\ 
G^{-+}(t,t^\prime) &= \vev{\phi^-(t) \phi^+(t^\prime)} \nn \\
G^{+-}(t,t^\prime) &= \vev{\phi^-(t^\prime) \phi^+(t)},
\label{eq:propzoo}
\end{align}
where we note that ``$-$'' fields are always to the left of ``$+$.''  The state indicated by the brackets, $\vev{\ldots}$ is implicitly the free vacuum, $\ket{0}$. For reference we note that $G^{++}$ is just the familiar Feynman propagator, with its time arguments shifted into the complex plane by $\eps$-deformation.  We can take $\phi^+$ to have its time argument deformed to $t \rightarrow t(1-i\eps)$ and $\phi^-$ to have $t \rightarrow t(1+i\eps)$.  Working in $t\textendash \vec k$ space familiar from cosmology, for a scalar with $\omega_k = \sqrt{\vec k^2 + m^2}$, this gives
\begin{align}
\phi^+_{\vec k}(t) &= \frac{1}{(2\pi)^{3/2} \sqrt{2\omega_k}} \left[ e^{-i\omega_k t} e^{-\eps t} \, a_{\vec k} \,+\,  e^{i\omega_k t}  e^{\eps t} \, a^\dag_{\vec k} \right] \nn \\
\phi^-_{\vec k}(t) &= \frac{1}{(2\pi)^{3/2} \sqrt{2\omega_k}} \left[ e^{-i\omega_k t} e^{\eps t} \, a_{\vec k} \,+\,  e^{i\omega_k t}  e^{-\eps t} \, a^\dag_{\vec k} \right],
\label{eq:phidefinit}
\end{align}
and we note that the $\phi^{\pm}$ are no longer hermitian.  Whenever a time will explicitly have a finite value, like the correlation time $\bar{t}$, we can set $\eps = 0$, similarly for $\eps$ appearing in any fixed power of $t$.  In practice, it is only the combination of $t \rightarrow -\infty$ and $t$ appearing in an exponential that require careful tracking of $\eps$.

Obtaining the $\epsilon$-deformed Green's functions that properly give projections of the free vacuum onto the interacting one for both bra and ket states is just a straightforward combination of \Eqs{eq:propzoo}{eq:phidefinit},
\begin{align}
G^{++}(t,t^\prime; \vec k) &= \frac{e^{-i \omega |t-t^\prime|} e^{-\eps |t-t^\prime|}}{2\omega_k} \nn \\ 
G^{--}(t,t^\prime; \vec k) &= \frac{e^{i \omega |t-t^\prime|} e^{-\eps |t-t^\prime|}}{2\omega_k} \nn \\ 
G^{-+}(t,t^\prime; \vec k) &= \frac{e^{-i \omega (t-t^\prime)} e^{\eps (t+t^\prime)}}{2\omega_k} \nn \\
G^{+-}(t,t^\prime; \vec k) &= \frac{e^{-i \omega (t^\prime-t)} e^{\eps (t+t^\prime)}}{2\omega_k}.
\label{eq:epszoo}
\end{align}
With these functions in hand, one can commence computing perturbatively.  

Ref.~\cite{Kaya:2018jdo} conjectured that a different, unitary $\eps$-deformation is possible, but which leads to identical perturbative correlation functions, $\bra{\Omega} \mo \ket{\Omega}(t)$.  We can arrive at this by the manifestly hermitian deformation of Eq.~\ref{eq:phidefinit}. One can think of this as just making the interactions time-dependent so as to vanish at asymptotically early times. The related $\eps$-deformed time evolution operator,  
\begin{equation}
U_{\eps} = T e^{ - i \int^{\bar{t}}_{-\infty} H_I e^{\eps t} dt},
\label{eq:udeform}
\end{equation}
remains exactly unitary. The derivation we gave in Ref.~\cite{Baumgart:2019clc} straightforwardly applies to this deformed Hamiltonian and results in the perturbative equivalence of
$\langle 0 | U_{\eps}^{\dagger} {\cal O} U_{\eps} | 0 \rangle$ with the nested commutator form, Eq.~\ref{eq:pertexprime}.  

The perturbative expansion of $\langle 0 | U_{\eps}^{\dagger} {\cal O} U_{\eps} | 0 \rangle$ involves fields $\phi^+$ and $\phi^-$  in the interactions from $U_{\eps}$ and $U^{\dagger}_{\eps}$ respectively, replacing  Eq.~\ref{eq:phidefinit}, 
\begin{align}
\phi^+_{\vec k}(t) &= \frac{e^{\epsilon t} }{(2\pi)^{3/2} \sqrt{2\omega_k}} \left[ e^{-i\omega_k t}  \, a_{\vec k} \,+\,  e^{i\omega_k t}   \, a^\dag_{\vec k} \right] \nn \\
\phi^-_{\vec k}(t) &= \frac{e^{\epsilon t}}{(2\pi)^{3/2} \sqrt{2\omega_k}} \left[ e^{-i\omega_k t}  \, a_{\vec k} \,+\,  e^{i\omega_k t}   \, a^\dag_{\vec k} \right], \nn \\
\Rightarrow \phi^+_{\vec k}(t) &= \phi^-_{\vec k}(t) \,=\, e^{\epsilon t} \phi_{\vec k}(t),
\label{eq:phidefinitalt}
\end{align}
where $\phi$ is the naive, undeformed scalar field, and thus $\phi^\pm$ are clearly hermitian.  To see that Eq.~\ref{eq:phidefinitalt} follows from Eq.~\ref{eq:udeform}, consider $\lambda \phi^4$ interactions, where we can trivially replace the $e^{\eps t}$ factor in Eq.~\ref{eq:udeform} by $e^{4 \eps t}$ since this 
is just rescaling the definition of $\eps$. One can then associate an $e^{\eps t}$ factor with each of the interacting fields $\phi$, as above. Similarly, fields inside correlator operators ${\cal O}$ can be multiplied by  $e^{\eps \bar{t}}$, since this will trivially approach $1$ as $\eps \rightarrow 0$ at the end of any calculation. Thus, these deformed fields  lead simply to new propagators,
\begin{align}
G^{++}_U(t,t^\prime; \vec k) &= \frac{e^{-i \omega |t-t^\prime|} e^{\eps (t+t^\prime)}}{2\omega_k} \nn \\ 
G^{--}_U(t,t^\prime; \vec k) &= \frac{e^{i \omega |t-t^\prime|} e^{\eps (t+t^\prime)}}{2\omega_k} \nn \\ 
G^{-+}_U(t,t^\prime; \vec k) &= \frac{e^{-i \omega (t-t^\prime)} e^{\eps (t+t^\prime)}}{2\omega_k} \nn \\
G^{+-}_U(t,t^\prime; \vec k) &= \frac{e^{-i \omega (t^\prime-t)} e^{\eps (t+t^\prime)}}{2\omega_k},
\label{eq:epszoou}
\end{align}
replacing Eq.~\ref{eq:epszoo}.  The subscript ``$U$" denotes ``unitary", in that they arise from the manifestly unitary $U_{\eps}$. 

 It is broadly
plausible that such an $\epsilon$ deformation might be equivalent to the original one following from Eq.~\ref{eq:inindef}.  After all, final physical answers have $\epsilon$ set to zero. In practice, we see that the action of the deformation to the propagators in \Eq{eq:epszoo} is minimal, if crucial.  When we perform the time integrals in computing correlation functions, the nonzero $\eps$ removes any ambiguity in the result as $t \rightarrow -\infty$.  In fact, it allows us to drop the contribution from the infinite past.  A healthy theory clearly does not admit any ambiguous terms, and yet we see that the role played by the $G^{\pm\pm}$ and $G^{\pm \mp}$ of Eq.~\ref{eq:epszoo} is very different.  In the latter, all early time contributions are exponentially suppressed.  In the former, though, we see that even if $t,\, t^\prime \rightarrow -\infty$, suppression does not automatically occur in the region $t-t^\prime \sim \; $const.  
This subtlety was overlooked in \cite{Kaya:2018jdo}.  

Tackling the $t-t^\prime \sim \; $const.~issue properly, we will show that we get an identical set of correlation functions whether we use the propagators of Eq.~\ref{eq:epszoo} or Eq.~\ref{eq:epszoou}, after $\eps \rightarrow 0$ at the end of calculations. And yet, since the latter are equivalent to the nested commutator reformulation of Eq.~\ref{eq:pertexprime}, so are the former, namely those of Eq.~\ref{eq:inindef}. Note that the analog of the denominator of Eq.~\ref{eq:inindef} is trivial in the unitary $\eps$ deformation, since $\langle 0 | U_{\eps}^{\dagger}  U_{\eps} | 0 \rangle = 1$ by unitarity and the normalization of $|0 \rangle$. Therefore we are really comparing the perturbative expansion of Eq.~\ref{eq:inindef} constructed using the $G$ propagators of Eq.~\ref{eq:epszoo} with the same diagrams constructed using the $G_U$ propagators of Eq.~\ref{eq:epszoou}, and checking they are the same after $\eps \rightarrow 0$. If so, then we have the equivalence of Eq.~\ref{eq:inindef} and Eq.~\ref{eq:pertexprime}.

Consider a generic in-in diagram following from the use of propagators, $G$, from Eq.~\ref{eq:epszoo}, and decompose it into partial contributions where the vertices are time-ordered.  Thus, the structure of the integral will be as follows,
\begin{align}
\vev{\mo}(\bar{t}) &\supset \, \lim_{\eps \rightarrow 0}  ~ \lim_{T \rightarrow - \infty}  \int \frac{d^3 k_1}{(2\pi)^3} \ldots f\left[k_1,\ldots \right] 
\nn \\
&\times \int_{T}^{\bar{t}} dt_V \, e^{i \omega^\eps_V t_V} \ldots \int_{t_{2}}^t dt_1  \, e^{i \omega^\eps_1 t_1} \, e^{i \omega_0 \bar{t}},
\label{eq:intschem}
\end{align}
where the use of ``$\supset$'' just indicates that this a particular perturbative in-in contribution with $V$ vertices and may represent just one of many time orderings.  We note that the function, $f$, captures  all of the pure momentum dependence (outside of energy-dependent phase factors) as well as any numerical factors.  The $\omega_j^\eps$ factors in the phases are linear combinations of (momentum-dependent) energies, whose overall sign may be positive or negative.  Before taking the limit, the RHS depends on $\eps$ within the slightly complexified  $\omega^\eps_i$, which we will discuss more explicitly below.  The earliest time limit of integration,  $T$, is explicitly written as limiting to $- \infty$.

Because all the time integrals involve exponential integrands, they are easily performed to give the form, 
\begin{align}
\vev{\mo}(\bar{t}) &\supset \, \lim_{\eps \rightarrow 0}  ~ \lim_{T \rightarrow - \infty}  
\int \frac{d^3 k_1}{(2\pi)^3} \ldots f\left[k_1,\ldots \right]  
\nn \\
&\quad \times  \sum_{i=0}^V R_i[\omega_1, ....] \, e^{ i \sum_{m=0}^i \omega_m \bar{t}} \, e^{ i \sum_{n=i+1}^V \omega_{n}^{\epsilon} T},
\label{eq:intschemev}
\end{align}
The $\omega_0$ factor always necessarily multiplies $\bar t$, since that combination appears from the original integrand.  The sum is over the partitions of the vertices to account for the different choice of integration limits for each time integral.  The fact that the division into $\bar t$ and $T$ terms has such a simple structure is due to the integrand's time-ordering.  $R_i$ is a rational function of the energy linear combinations, $\omega_j$, arising from doing the integrals. Note, we have neglected $\epsilon$ everywhere except in the exponentials involving the earliest time $T$, since this tends to $-\infty$. After performing all time integrals, the only remnant of each vertex time $t_j$ is its possible limiting values of $\bar{t}$ and $T$. We can now study the origin and fate of the $\epsilon$ dependence. 

The $\epsilon$ dependence originates from the propagators in   Eq.~\ref{eq:epszoo}, appearing as small imaginary corrections to the energies $\omega$ that appear there. Linear combinations of such $\epsilon$-corrected $\omega$ constitute the $\omega_j^{\epsilon}$ above.  While in the $G^{+-}$ and $G^{-+}$ propagators, this dependence always gives a damping factor for any sufficiently early (negative) time $t$,  $e^{\epsilon t} < 1$, this is clearly not always the case in the  $G^{++}$ and $G^{--}$ propagators, where it is possible for a time $t$ to have a  $e^{- \epsilon t}$ factor which is an enhancement for early time. The burning question is then whether the early time factor in Eq.~\ref{eq:intschemev} above, 
$e^{ i \sum_{n=i+1}^V \omega_{n}^{\epsilon} T}$, is (a) damped by the accumulated $\epsilon$ factors, in which case it is simply set to zero after the $T \rightarrow - \infty$ limit, or (b)  has  a net enhancement factor or complete cancelation of $\epsilon$ factors so that the $T \rightarrow - \infty$ limit is ill-defined. We will show that except for the trivial case where $i=V$, and thus all vertex times are evaluated at $\bar t$, making the early time factor of Eq.~\ref{eq:intschemev} just $1$, (a) is always the case, and there is therefore a well-defined $T \rightarrow - \infty$ limit in which the early time factor vanishes. Consequently, after the $T \rightarrow - \infty$ limit, Eq.~\ref{eq:intschemev} collapses to
\begin{align}
\vev{\mo}(\bar{t}) &\supset \, 
\int \frac{d^3 k_1}{(2\pi)^3} \ldots f\left[k_1,\ldots \right]  R_V[\omega_1, ....]
\nn \\
&\quad \times   e^{ i \sum_{m = 0}^V \omega_m \bar{t}}.
\label{eq:intschemwd}
\end{align}

\begin{figure}[ht!]
\begin{center}
\includegraphics[width=9cm]{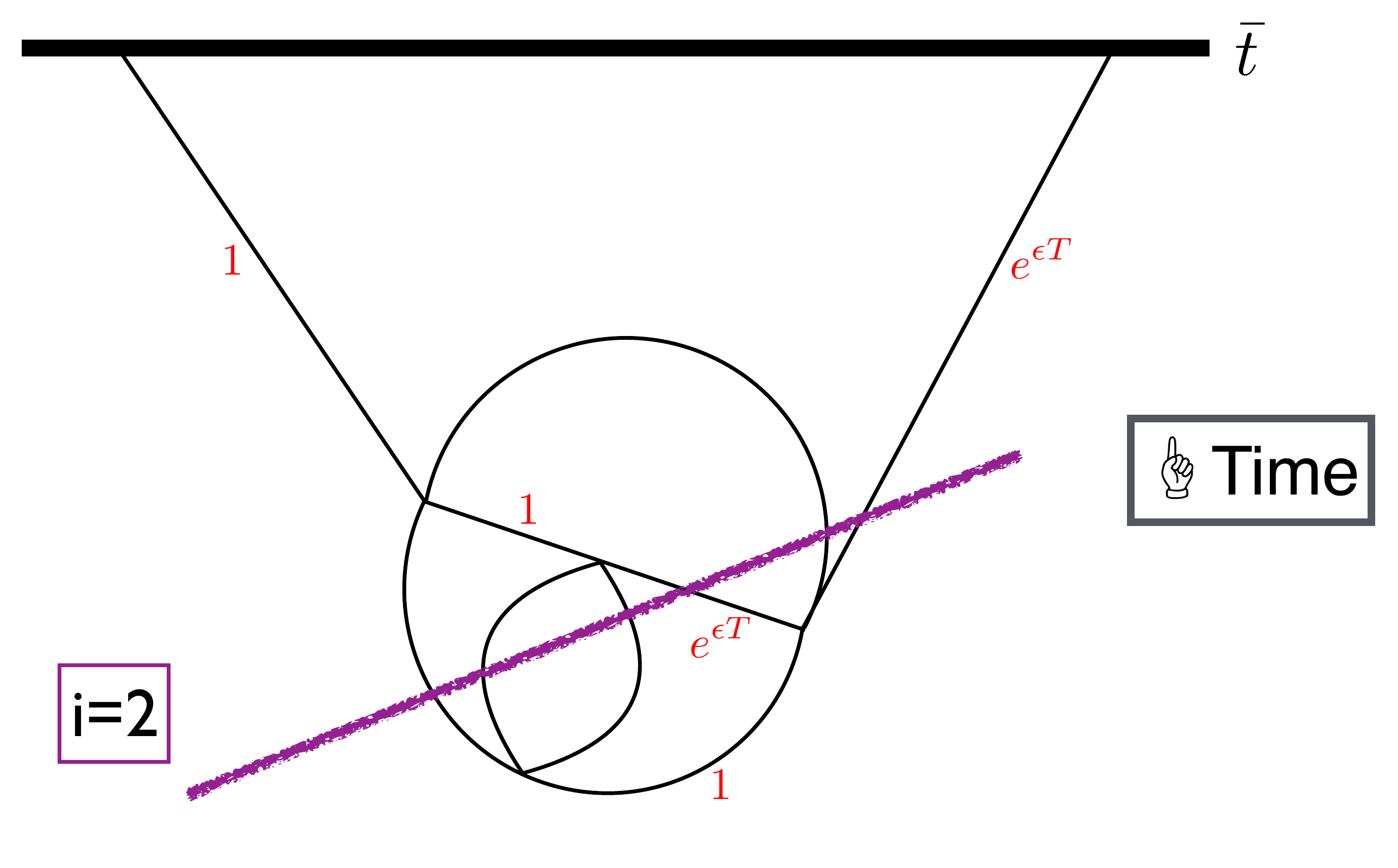}
\caption{An in-in diagram with all $G^{\pm\pm}$ propagators where we demonstrate the structure of contributions to Eq.~\ref{eq:intschemev}.  The coarse, purple line is a partition that divides vertices into those whose time argument evaluates to the final correlation time, $\bar t$, from those set to  the early-time cutoff, $T$.  Since the original integral (Eq.~\ref{eq:intschem}) is time-ordered, this dividing line occurs between the $i^{\rm th}$ and $(i+1)^{\rm th}$ vertices.  Here we show it in a four vertex graph for $i=2$.  
The red labels on some propagators show the $\eps$ dependence, 1 or $e^{\eps T}$, that results from deformation ({\it cf.}~Eq.~\ref{eq:gsets}).  We see that the factor is 1 if the propagator connects two vertices on the same side of the partition.  If, however, a propagator straddles the partition, then there is a suppressing $e^{\eps T}$.  Furthermore, because the propagators that connect to the external correlation point have one end evaluated at $\bar t$ even before integration, the external correlation points are always above the partition line; the only way to avoid suppression is if every vertex evaluates to $\bar t$ (corresponding to the partition drawn below the {\it entire} graph and cutting no propagators), which is precisely the surviving contribution in Eq.~\ref{eq:intschemwd}.  We note that this straddler argument fails for the case of a vacuum bubble, as all vertices can evaluate to $T$ without any suppression, giving a finite contribution.
}
\label{fig:early}
\end{center}
\end{figure}

To show that the ill-defined case of (b) does not arise, we proceed as follows. We first note that after doing all $t_j$ time integrals, in any contribution these times are evaluated at one of the integration limits, $\bar{t}$ or $T$. Therefore every propagator between any $t_j$ and $t_l$, ends up contributing possible $\epsilon$ dependence obtained from Eq.~\ref{eq:epszoo} by choosing $t_j$ and $t_l$ from either $\bar{t}$ or $T$. The $\epsilon$ dependences are therefore given by propagators chosen from the following sets,
\begin{align}
\big \{ G^{++}(\bar{t}, T),  G^{++}(T,\bar{t}), G^{--}(\bar{t}, T), G^{--}(T,\bar{t}) \big \} &\propto e^{- \epsilon |\bar{t} - T|} \approx e^{ \epsilon T}   \nn \\ 
\big \{ G^{++}(\bar{t}, \bar{t}),  G^{++}(T,T), G^{--}(\bar{t}, \bar{t}), G^{--}(T,T) \big \} &\propto 1 \nn \\ 
\big \{ G^{+-}(\bar{t}, T),  G^{+-}(T,\bar{t}), G^{-+}(\bar{t}, T), G^{-+}(T,\bar{t}) \big \} &\propto e^{\epsilon \bar{t} + T} \approx e^{ \epsilon T}   \nn \\ 
\big \{ G^{+-}(\bar{t}, \bar{t}),   G^{-+}(\bar{t}, \bar{t}) \big \} &\propto e^{ 2 \epsilon \bar{t} } \approx 1  \nn \\ 
\big \{ G^{+-}(T,T),   G^{-+}(T,T) \big \} &\propto e^{2 \epsilon  T} \approx e^{\epsilon T},
\label{eq:gsets}
\end{align}
from which we see that there are  no exponential enhancements as $T \rightarrow - \infty$. There are at worst the $1$ terms on the second line where vertex times are evaluated at $T$ and yet there is no exponential damping factor. It is then easy to see that 
 in any diagram that connects to the correlator time $\bar{t}$, 
there is always a contribution to the early time factor of Eq.~\ref{eq:intschemev} of our type (a); the terms from Eq.~\ref{eq:gsets} cannot come exclusively from (b). This is because for any $i<V$ there is  at least one $t_j$ vertex time which is evaluated at $T$, so there must then be some propagator in this diagram which straddles from time $T$ to correlator time $\bar{t}$, and is therefore $\propto e^{\epsilon T}$ ({\it cf.}~Fig.~\ref{fig:early}). Given that no other propagator has an exponential enhancement that can cancel this, as seen above, a net suppression must remain, corresponding  to our case (a). 

If instead we had used the alternate $G_U$ propagators of Eq.~\ref{eq:epszoou}, the analysis would be much easier because their $\epsilon$ dependence is universally $\propto e^{\epsilon (t + t^\prime)}$. Consequently, any contribution in Eq.~\ref{eq:intschemev} with $i<V$ would have the early time factor $\propto e^{\epsilon N T}$ for some strictly positive integer $N$, and would therefore vanish as $T \rightarrow - \infty$. Thus, again Eq.~\ref{eq:intschemev} would collapse to Eq.~\ref{eq:intschemwd}.  This completes the proof of the equivalence of perturbative diagrams that connect to the correlator time $\bar{t}$ whether we use propagators $G$ from Eq.~\ref{eq:epszoo} or $G_U$ from Eq.~\ref{eq:epszoou}.

There is one remaining subtlety, missed in Ref.~\cite{Kaya:2018jdo}, namely diagrams
that do not connect to the correlator time $\bar{t}$. For such graphs we cannot show that building them from the propagators of Eq.~\ref{eq:epszoo} always gives early time 
factors that are suppressed by $e^{\epsilon N T},  ~ N > 0$ because we clearly lack the ``straddling" argument above. This conflicts with the use of the $G_U$ propagators 
from Eq.~\ref{eq:epszoou}, which obviously yield such a suppression. But such diagrams are precisely the 
``vacuum bubbles" that are canceled in taking the ratio of Eq.~\ref{eq:inindef}. Thus, the net diagrammatic expansion following from Eq.~\ref{eq:inindef} for the in-in correlator is indeed the same
whether computed by using $G$ of Eq.~\ref{eq:epszoo} or $G_U$ of Eq.~\ref{eq:epszoou}. However, as discussed when introducing $G_U$, the diagrams that follow from its use are the 
perturbative expansion of the in-in correlators using the $\epsilon$-deformation of Eq.~\ref{eq:udeform}, which maintains manifest exact unitarity of the $\eps$-deformed 
time evolution operators. This undeformed exact unitarity allows us to then derive the Weinberg formulation, Eq.~\ref{eq:pertexprime}, unimpeded. 

Assembling all the results for Minkowski spacetime, we have shown that the original in-in perturbation theory following from Eq.~\ref{eq:inindef}, with the standard $\eps$-deformed propagators of Eq.~\ref{eq:epszoo} is equivalent to Weinberg's reformulation in time-dependent perturbation theory, Eq.~\ref{eq:pertexprime}.

\vspace{0.3in}

\section{De Sitter}
\label{sec:ds}

For our De Sitter (DS) derivation, we start in FRW coordinates, and change the time variable  from FRW proper time to conformal time, $\eta\; (= -e^{-Ht}/H$).  This gives the following form for the metric,
\beq
ds^2 = \frac{1}{(H\eta)^2} \left( d\eta^2 - d\vec x^2  \right).
\eeq
Here, $H$ is the constant Hubble scale.
The construction of a manifestly unitary and causal in-in formalism for De Sitter follows from a very similar argument to flat space.  The modifications are simply technical, as propagators are given at asymptotically early times (where the details of different $\eps$ prescriptions matter) 
by asymptotic series of terms $\sim e^{-ik \eta}/(k \eta)^n$, rather than the simple exponentials of time we dealt with in Sec.~\ref{sec:mink} ({\it cf.}~Eq.~\ref{eq:epszoou}).  This follows from the exact solution of the De Sitter equations of motion.  Taking the free theory ground state to be the Bunch-Davies vacuum, we have the following expression for a scalar field,
\beq
\phi_{\vec k}(\eta) = \frac{H}{\sqrt{2} \, (2\pi)^{3/2}} \left[ \eta^{3/2} H^{(2)}_\nu (k \eta) \, a_{\vec k} \,+\,  \eta^{3/2} H^{(1)}_\nu (k \eta) \, \, a^\dag_{\vec k} \right],
\label{eq:scds}
\eeq
where $H^{(1,2)}_\nu$ are Hankel functions with $\nu = \sqrt{9/4 - m^2/H^2}$ in 4D \cite{Allen:1985ux}.\footnote{In $D$ dimensions this switches to $\nu = \sqrt{(D-1)^2/4 - m^2/H^2}$.}  

Much of the technical challenge of DS arises from the fact that $H^{(1,2)}_\nu$ are not elementary functions unless $\nu$ is a half odd integer.  In fact, for the exactly massless case, $\nu = 3/2$.  However, the steps involved in validating manifestly unitary and causal in-in perturbation theory for $m=0$ are just a subset of those needed for general $m$.  We therefore proceed with the analysis for arbitrary $\nu \in \mathbb{R}$.  At early times (large $|\eta|$) where the $\eps$ prescription is relevant, $H^{(1,2)}_\nu$ admit asymptotic expansions of the form, 
\begin{align}
H^{(1)}_\nu(z) &\sim \left( \frac{2}{\pi \, z} \right)^{1/2} e^{i(z - \frac 1 2 \nu \, \pi - \frac 1 4 \pi)} \sum_{j=0}^{\infty} i^j \frac{a_j(\nu)}{z^j} \nn \\
H^{(2)}_\nu(z) &\sim \left( \frac{2}{\pi \, z} \right)^{1/2} e^{-i(z - \frac 1 2 \nu \, \pi - \frac 1 4 \pi)} \sum_{j=0}^{\infty} (-i)^j \frac{a_j(\nu)}{z^j},
\label{eq:hexp}
\end{align}
in the sense that if one truncates the series at the $j= N$ term, the corrections scale as $|z|^{-(N+1)}$  for large $|z|$ \cite{nemes}.\footnote{The series is exact and contains only a finite number of terms if $\nu$ is a half-odd integer, which is the case for an exactly massless particle.}  The presence of the $z^{-1/2}$ prefactor generally places a branch cut on the negative real axis.  However, in our physical case of interest, propagators are constructed by products of the mode functions, $\eta^{3/2} H^{(1,2)}_\nu(k \eta) $, giving strictly integer powers  The numerator factors, $a_j(\nu)$, can be found in any standard reference ({\it e.g.}~DLMF, Section 10)\cite{dlmf}, but we do not require their detailed form.  


In diagrams, the Hankel functions appear via the four Green's functions that make up our initial basis for in-in perturbation theory, $G^{\pm\pm},\, G^{\pm\mp}$, which are still defined by Eq.~\ref{eq:propzoo}.  In the standard in-in DS perturbation theory, we deform the expression for $\phi$ given in Eq.~\ref{eq:scds} to get
\begin{align}
\phi^+_{\vec k}(\eta) &= \frac{H}{\sqrt{2} \, (2\pi)^{3/2}} \left[  \eta^{3/2} H^{(2)}_\nu \big[ k \eta (1 - i\eps) \big] \, a_{\vec k} \right. \nn \\
   & \quad \left. + \eta^{3/2} H^{(1)}_\nu \big[ k \eta (1 - i\eps) \big] \, \, a^\dag_{\vec k} \right], \nn \\
\phi^-_{\vec k}(\eta) &= \frac{H}{\sqrt{2} \, (2\pi)^{3/2}} \left[  \eta^{3/2} H^{(2)}_\nu \big[ k \eta (1 + i\eps) \big] \, a_{\vec k} \right. \nn \\
   & \quad \left. + \eta^{3/2} H^{(1)}_\nu \big[ k \eta (1 + i\eps) \big] \, \, a^\dag_{\vec k} \right],
\label{eq:scdsdef}
\end{align}
for fields arising from the ket and bra evolution, respectively.  In the early-time regime, large $|\eta|$, where the $\eps$ prescription matters, it is sufficient to keep the $\eps$ in just the exponential factors of  the asymptotic Hankel expansions of Eq.~\ref{eq:hexp}, so that we can factor out the $\eps$ deformation of the mode functions just as we did in flat space,
\begin{align}
\phi^+_{\vec k}(\eta) &\rightarrow \frac{H}{\sqrt{2} \, (2\pi)^{3/2}} \left[ e^{-\eps \eta} \, \eta^{3/2} H^{(2)}_\nu (k \eta ) \, a_{\vec k} \right. \nn \\
   & \quad \left. + e^{\eps \eta} \, \eta^{3/2} H^{(1)}_\nu (k \eta ) \, \, a^\dag_{\vec k} \right], \nn \\
\phi^-_{\vec k}(\eta) &\rightarrow \frac{H}{\sqrt{2} \, (2\pi)^{3/2}} \left[ e^{\eps \eta} \, \eta^{3/2} H^{(2)}_\nu (k \eta ) \, a_{\vec k} \right. \nn \\
   & \quad \left. + e^{-\eps \eta} \, \eta^{3/2} H^{(1)}_\nu ( k \eta ) \, \, a^\dag_{\vec k} \right].
\label{eq:scdsdefexp}
\end{align}
This exponential $\eps$ dependence is stronger than the $\eps$ dependence in the powers of $\eta$ in the Hankel expansions, and 
 will be shown below to always give overall exponential suppression for early $\eta$ in diagrammatic integrals, as was the case in flat space. This 
  justifies dropping the subleading $\eps$ dependence in the powers of $\eta$. 
  We use 
 ``$\rightarrow$'' as this simple $\eps$-deformation structure only strictly holds for finite $\eps$ at early times, but it will result in the same value of diagrams as the defining $\eps$-deformation once we finally take $\eps \rightarrow 0$, since we are only sensitive to $\eps$ when integrating over early times. 
Thus, in what follows we study the $\eps$-deformed $\phi^\pm$ in Eq.~\ref{eq:scdsdefexp} at {\it all} times, as equivalent to the defining in-in $\eps$-deformation.

If we plug in the deformed $\phi^\pm$ fields from Eq.~\ref{eq:scdsdefexp} to get the in-in propagators, the $\eps$-dependence is given by the following overall factors for each,
\begin{align}
G^{++}(\eta,\eta^\prime; \vec k) &\propto e^{-\eps |\eta-\eta^\prime|} \nn \\ 
G^{--}(\eta,\eta^\prime; \vec k) &\propto e^{-\eps |\eta-\eta^\prime|} \nn \\ 
G^{-+}(\eta,\eta^\prime; \vec k) &\propto e^{\eps (\eta+\eta^\prime)} \nn \\
G^{+-}(\eta,\eta^\prime; \vec k) &\propto e^{\eps (\eta+\eta^\prime)}.
\label{eq:epszoods}
\end{align}
These are identical in form to the deformation factors in the flat space propagators (Eq.~\ref{eq:epszoo}) just by switching $\eta \rightarrow t$.  We showed in Section \ref{sec:mink} that the final result for a Minkowski correlation function, $\bra{\Omega} \mo \ket{\Omega}(t)$, is identical if we switch to the $G_U$ propagators (Eq.~\ref{eq:epszoou}).  The question for the De Sitter case is whether the analogous DS $G_U$ propagators, defined by the alternative $\eps$ prescription,
\begin{equation}
G_U \propto e^{\eps (\eta+\eta^\prime)},
\end{equation}
gives an identical result to the $G$ in Eq.~\ref{eq:epszoods} once $\eps \rightarrow 0$ in diagrams. The central difference is that in De Sitter we do not have purely exponential integrals, even at asymptotically early times, as the asymptotically-early integrands generally also have (negative) powers of $\eta$ from Eq.~\ref{eq:hexp}, as well as from measure factors for the metric. There may also be positive powers from inverse metric factors if there are derivative interactions, and higher-point non-derivative interactions.

To prove equivalence of the different propagator deformations, we again write down a general time-ordered perturbative contribution to an in-in correlator starting with the  effectively-standard $G$ propagators in Eq.~\ref{eq:epszoods}: 
\begin{align}
\vev{\mo}(\bar \eta) &\supset \, \lim_{\eps \rightarrow 0} \,\lim_{T \rightarrow - \infty} \int \frac{d^3 k_1}{(2\pi)^3} \ldots f\left[k_1,\ldots \right]  
\nn \\
&\times  \int_{T}^{\bar \eta} d\eta_V \, F_V(\eta_V) \int_{\eta_{V}}^{\bar \eta} d\eta_{V-1} \, F_{V-1}(\eta_{V-1}) 
 \ldots  \nn \\
 &\ldots \int_{\eta_2}^{\bar \eta} d\eta_{1}  F_{1}(\eta_{1}) ~F_{0}(\bar \eta),
\label{eq:dsintschem}
\end{align}
Here, each $F_j$ involves products of Hankel functions (and other metric and measure dependent powers of $\eta$), and therefore has an early-time asymptotic expansion of the form,
\begin{equation}
F_j(\eta_j) \sim \sum_n b_n^{(j)}\,  \frac{e^{i \omega^\eps_j \eta_j}}{\eta_j^{n}}.
\end{equation}
%
The $\omega$ depend on the $k$, and as in Minkowski space, they depend on $\eps$ via Eq. Eq.~\ref{eq:epszoods}.  We have kept the same notation for the early time cutoff, $T$, as in flat space, even though here $T$ is a conformal time. 

Fortunately, we do not need to explicitly do these integrals, but only sequentially ascertain the form of their asymptotic expansions at their early limit of integration, based on the asymptotic expansion of their integrands. The key to this is 
given by the asymptotic form
\begin{equation}
\int_{\eta}^{\bar \eta} d \eta' F(\eta') \sim c(\bar \eta) + \sum_n c_n\, \frac{e^{i \omega^{\eps} \eta}}{\eta^{n}},
\label{eq:asympint}
\end{equation}
if $F$ has an asymptotic expansion for large $\eta^\prime$,
\begin{equation}
F(\eta') \sim \sum_n d_n\, \frac{e^{i \omega^{\eps} \eta'}}{\eta^{\prime n}}.
\label{eq:asympform}
\end{equation}
This follows because the asymptotic dependence on $\eta$ depends on the integration for early time $\eta'$, where we can use the integrand's asymptotic expansion.
The integral of each term is given by an  incomplete gamma function because of the identity,
\beq
 \frac{d}{dx} (-i)^{n+1} \Gamma[1-n,-ix] = \frac{e^{i\, x}}{x^n}.
\label{eq:intofi}
\eeq
%
We note that $\Gamma[a,b]$ is holomorphic in the vicinity of the imaginary axis away from zero, which is our domain of  interest given $x \equiv \omega^{\eps} \eta'$ ({\it cf.}~DLMF, Section 8 \cite{dlmf}).  This then leads to Eq.~\ref{eq:asympint}, given the large-$x$ asymptotics (Ref.~\cite{olver}, Eq.~2.02), 
\beq
(-i)^{n+1} \Gamma[1-n,-ix] =\, \frac{e^{ix}}{x^n} \left[ -i + O \left( \frac{1}{x} \right)\right],
\label{eq:appofi}
\eeq
where the corrections in the square brackets are negative powers of $x$.  The $c(\bar \eta)$ in Eq.~\ref{eq:asympint} is an ``integration constant" with respect to the $\eta$ dependence. 

In this way, we see that starting with $\eta_1$ and ending with $\eta_V$, we repeatedly get integrals of the form of Eq.~\ref{eq:asympint} with the integrands having Eq.~\ref{eq:asympform} as their asymptotic early-time behavior. 
%
%
%
%
We thereby get, in analogy with Eq.~\ref{eq:intschemev} in Section \ref{sec:mink},  a sum of early-$T$ and $\eps$ dependent contributions of the form
\begin{align}
\propto e^{ i \sum_{a=i}^V \omega_{a}^{\epsilon} T} \frac{1}{T^{N_i}}.
\label{eq:fineval}
\end{align}
At each stage of integration, we must sum over the choice to follow either the ``integration constant" $c(\bar \eta)$ or the asymptotic series in Eq.~\ref{eq:asympint}. The index $i$ in Eq.~\ref{eq:fineval} represents the largest index in any particular contribution for which the ``integration constant" option is chosen for the asymptotics of the $\eta_{i-1}$ integral (in the notation of the diagram in Fig.~\ref{fig:early}, the contribution of Eq.~\ref{eq:fineval} puts vertices 1 through $i-1$ above the partition line).


The logic from here on is identical to the Minkowski case, given the nearly identical structure of Eq.~\ref{eq:epszoods} (with $t \rightarrow \eta$).  Again, Eq.~\ref{eq:epszoods} implies that there cannot be exponential enhancements in $T$ coming from $\eps$ deformation, only exponential damping or neutral independence of $\eps$.  Also paralleling the Minkowski case, we see from Eq.~\ref{eq:fineval} that there are some early vertices $\eta_i, \ldots \eta_V$ evaluated at $T$, while the later times $\eta_1, \ldots, \eta_{i-1}$ only get $T$-independent contributions. If the entire diagram connects to correlator time $\bar \eta$, there must be straddling propagators from the early vertices to the later ones, implying at least one suppression factor in $T$, $e^{\eps T}$, from such a propagator, and there is no possibility of cancelling this suppression with exponential enhancement in $T$.  
 Therefore all such graphs where any vertex time integral is evaluated at $T$ have net exponential damping, and therefore vanish as $T \rightarrow - \infty$. Dropping all such contributions of the form of Eq.~\ref{eq:fineval}  is the entire action of the standard in-in $\eps$ prescription, after which we can safely take $\eps \rightarrow 0$.  
 
The ``straddling" argument again fails  for graphs which are not connected to correlator time $\bar{\eta}$, but these ``vacuum bubble" graphs are canceled in the ratio of Eq.~\ref{eq:inindef}. Just as for flat space, if we switch to the $G_U$ propagators of the unitary $\eps$ deformation, which all have $e^{\eps(\eta + \eta^\prime)}$ suppression, these straightforwardly eliminate precisely any contributions of the form of Eq. (29),  just as we non-trivially found for the standard $G$ propagators.  There are also again no vacuum bubble graphs with the $G_U$ propagators, as they vanish by the preserved unitarity of time evolution ({\it cf.}~Eq.~\ref{eq:udeform}, which has a trivial DS equivalent). 
 
 This then proves that standard in-in
  perturbation theory in De Sitter is equivalent to the manifestly unitary reformulation, and therefore Eq.~\ref{eq:pertexprime} follows, by which it is manifestly causal, as well.

\vspace{0.3in}

{\it Acknowledgments:} 
We thank Ali Kaya for useful discussions and Gerg\H{o} Nemes for correcting our understanding of the asymptotic expansions.  We also acknowledge the anonymous JHEP referee of our previous paper for bringing the issue addressed here to our attention.  MB is supported by the U.S. Department of Energy, under grant number DE-SC0019470. RS is supported by NSF grant PHY-1914731 and by the Maryland Center for Fundamental Physics (MCFP).

\end{document}